# Kilovolt-Class $\beta$-Ga$_2$O$_3$ Field-Plated Schottky Barrier Diodes with MOCVD-Grown Intentionally 10$^{15}$ cm$^{-3}$ Doped Drift Layers

Carl Peterson,[1,a)] Chinmoy Nath Saha,[1] Rachel Kahler,[1] Yizheng Liu,[1] Akhila Mattapalli,[1] Saurav Roy,[1] and Sriram Krishnamoorthy[1,a)]

[1]*Materials Department, University of California Santa Barbara, Santa Barbara, California, 93106, USA*

---

[a)] Author to whom correspondence should be addressed. Electronic mail: carlpeterson@ucsb.edu and sriramkrishnamoorthy@ucsb.edu

We report on the growth optimization of intentionally low-doped (10$^{15}$ cm$^{-3}$) high-quality $\beta$-Ga$_2$O$_3$ drift layers up to 10 $\mu$m thick via MOCVD and the fabrication of kilovolt-class field plated Schottky barrier diodes on these thick drift layers. Homoepitaxial growth was performed on (010) $\beta$-Ga$_2$O$_3$ substrates using TMGa as the Ga precursor. Growth parameters were systematically optimized to determine the best conditions for high quality thick growths with the given reactor geometry. Chamber pressure was found to improve the growth rate, mobility, and roughness of the samples. Growth rates of up to 7.2 $\mu$m/hr., thicknesses of up to 10 $\mu$m, Hall mobilities of up to 176 cm$^2$/Vs, RMS roughness down to 5.45 nm, UID concentrations as low as 2 x 10$^{15}$ cm$^{-3}$, and controllable intentional doping down to 3 x 10$^{15}$ cm$^{-3}$ were achieved. Field plated Schottky barrier diodes (FP-SBDs) were fabricated on a 6.5 x 10$^{15}$ cm$^{-3}$ intentionally doped 10 $\mu$m thick film to determine the electrical performance of the MOCVD-grown material. The FP-SBD was found to have current density >100 A/cm$^2$ at 3 V forward bias with a specific differential on resistance (R$_{on,sp}$) of 16.22 m$\Omega$.cm$^2$ and a turn on voltage of 1 V. The diodes were found to have high quality anode metal/semiconductor interfaces with an ideality factor of 1.04, close to unity. Diodes had a maximum breakdown voltage of 1.50 kV, leading to a punch-through maximum field of 2.04 MV/cm under the anode metal, which is a state-of-the-art result for SBDs on MOCVD-grown (010) drift layers.

# I. INTRODUCTION

In the past decade, the Ultra-Wide Band Gap (UWBG) semiconductor Beta Gallium Oxide ($\beta$-$Ga_2O_3$) has shown immense promise for creating more efficient power electronic devices[1]. A wide bandgap (WBG) allows semiconducting materials to sustain higher electric fields before breaking down and thus devices on WBG materials can be made thinner and less resistive. WBG semiconducting materials such as GaN and SiC have already demonstrated the inherent advantages of a WBG for efficient power devices[2]. Next-generation UWBG materials such as $\beta$-$Ga_2O_3$ can further push these limits to create electronic devices capable of holding many kilovolts while maintaining low on-resistances. With $\beta$-$Ga_2O_3$'s bandgap of 4.6-4.9 eV, predicted critical electric field strength of 8 MV/cm, and availability of shallow dopants, figure of merit analyses show $\beta$-$Ga_2O_3$ to be the best-in-class material among its WBG and UWBG peers[3,4]. $\beta$-$Ga_2O_3$ is also the only WBG/UWBG material to demonstrate melt-grown conductive and insulating bulk substrates with dopant impurity control[5–11], potentially enabling cheaper production costs, large area and low extended defect density substrate platforms, and reduced defect density for epitaxial growth.

## A. MOCVD

High quality epitaxial growth is vital for any semiconductor device material. A promising epitaxial method for $\beta$-$Ga_2O_3$ is metalorganic chemical vapor deposition (MOCVD) due to its scalable growth rates[12–23], high electron mobilities[12,14,23–28], material alloying[29,30], in-situ etching[31–33], in-situ dielectrics[34–36], wide range of n-type conductivity[37,38], and delta doping capabilities[39,40]. These features have all been utilized to create state-of-the-art electronic devices on MOCVD-grown material[41–52]. To realize the next generation of high voltage power devices, it is critical to be able to create thick (10-100 μm) epilayers with controlled low net doping ($N_D$-$N_A$ ~$10^{15}$ -$10^{16}$ cm$^{-3}$) and high carrier mobility. Recently, MOCVD-grown β-$Ga_2O_3$ drift layers have shown promise in accomplishing these goals[12,13,16,17,19,22,23], including epilayers with controllable doping down to 4 x $10^{15}$ cm$^{-3}$ and mobilities reaching the theoretical maximum of 200 cm$^2$/Vs for gallium oxide[23].

## B. TMGa Vs. TEGa

One main research topic for growing high quality MOCVD drift regions is whether to use triethylgallium (TEGa) or trimethylgallium (TMGa) as the Ga precursor. Previous work with TEGa as the Ga precursor demonstrates very high quality films with growth rates up to 4.5 μm/hr., low electron concentrations, mobilities reaching the theoretical max, and unintentional carbon or hydrogen below SIMS detection limit[12,23]. The high quality and low impurity concentration for TEGa grown films can be attributed to the β-hydride elimination process of the ethyl group, which effectively eliminates C and H impurities from incorporating during growth[53]. The tradeoff for TEGa is that it has a lower vapor pressure and slower reaction kinetics when compared to TMGa, so achieving faster growth rates for cost-effective thick growth can be an issue. TMGa, on the other hand, has demonstrated very high growth rates up to 16 μm/hr.[15,20] but has the potential for C and H incorporation[18,54,55] which can cause free carrier compensation and lower mobilities. In this work, we optimized TMGa MOCVD growth conditions to realize a 10 μm thick intentionally doped $6.5 \times 10^{15}$ $cm^{-3}$ film. We then fabricated a 1.50 kV field plated Schottky barrier diode on the thick MOCVD-grown drift layer which demonstrated a peak parallel plane electric field of 2.04 MV/cm, a state-of-the-art result.

## II. MOCVD Growth

To optimize thick epilayer growth conditions, a systematic growth parameter sweep was performed on the MOCVD reactor. Growth was performed on an Agnitron Agilis 100 cold-wall MOCVD reactor with a remote injection vertical showerhead. For this study, all growths were performed on semi-insulating (010) Fe-doped β-$Ga_2O_3$ substrates with no intentional miscut, purchased commercially from NCT Japan. Prior to growth, substrates were cleaned with Acetone, Methanol, and DI water. Substrates then underwent a 30 minute 49% HF acid treatment to remove any surface silicon contamination and were then immediately loaded into the MOCVD reactor in under 10 minutes to prevent silicon re-contamination[24,56]. All growths were performed using trimethylgallium (TMGa) as the Ga precursor, pure $O_2$ gas as the oxygen source, Ar as the carrier gas, and silane ($SiH_4$) as the silicon dopant source for intentionally doped films. An initial high $Ga_2O_3$ growth temperature and $O_2$ molar flow rate of 1000 °C and 0.089 mol/min. were selected, respectively. These conditions were chosen to reduce potential carbon and hydrogen impurities in the films as C and H incorporation has been demonstrated to be a strong function of temperature and VI/III ratio when using TMGa as the Ga precursor[55], with temperatures lower than 900 °C and VI/III ratios lower than 615 showing Carbon contamination[18,54].

### A. Growth rate Vs. TMGa flow and reactor pressure

With the $O_2$ flow rate (0.089 mol/min.), growth temperature (1000 °C), growth pressure (60 Torr), and growth time (1 hr.) fixed, the TMGa flow was varied from 20 – 340 μmol/min, leading to a linear growth rate increase from 0.55 – 4.5 μm/hr. as seen in Figure 1(a). Growth thickness (and therefore growth rate) was determined by co-loading a c-plane sapphire substrate along with the $Ga_2O_3$. Since the growth rate of $Ga_2O_3$ on sapphire is nearly identical to that of homoepitaxial films, the co-loaded sapphire wafers were cleaved and the thickness of the heteroepitaxial $Ga_2O_3$ epilayer was measured via scanning electron microscopy (SEM). At a fixed TMGa flow, decreasing the growth pressure was found to increase the growth rate. For example, Figure 1(b) shows that at a fixed TMGa flow of 170 μmol/min., the growth rate increased from 2.7 – 4.2 μm/hr. when decreasing the pressure from 60 Torr to 15 Torr, with a maximum growth rate of 7.2 μm/hr. also achieved at 15 Torr and higher TMGa flows. The growth rate increasing with lower chamber pressure is most likely due to parasitic gas-phase pre-reactions. For the remote injection showerhead reactor geometry, decreasing the growth chamber pressure will increase the gas phase velocity, leading to a decrease in the residence time of TMGa in the growth chamber[57]. Since TMGa is reactive with oxygen in the gas phase, decreasing the TMGa residence time reduces the amount of parasitic gas-phase pre-reactions between TMGa and $O_2$, enabling more TMGa to reach the growth interface. Thus, reducing the pressure improves TMGa consumption efficiency and increases the growth rate. This trend of increasing growth rate with a reduction in pressure has been seen in other $Ga_2O_3$ works using both TEGa and TMGa as the Ga precursor[12,58–60].

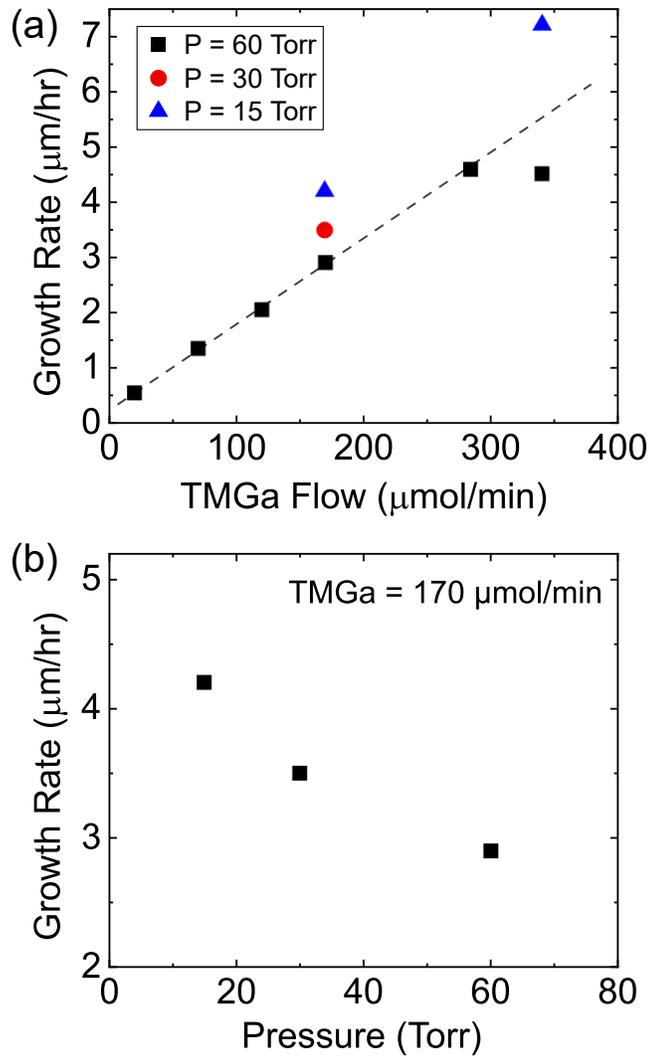

FIG. 1. (a) Plot of TMGa flow rate vs. β-Ga$_2$O$_3$ growth rate and (b) pressure vs. β-Ga$_2$O$_3$ growth rate, showing that growth rate increases with TMGa flow rate and decreases with increased pressure.

## B. Hall measurement and surface morphology

The room temperature electron transport properties of the film were studied via Van Der Pauw Hall structures. For these structures, Ti/Au contacts were deposited on the 4 corners of a square 5x5 mm$^2$ sample via electron beam (e-beam) evaporation to create ohmic contacts. 4-corner Hall contacts are not exact, and can affect the measurement by up to 20% error based on the size and uniformity of the corner contacts[61], for more accurate results, lithographically defined mesa-isolated Hall structures can be used, as seen in the supplementary information of reference 23. Nevertheless, the fabricated 4-corner Hall structures give a good estimate of the epitaxial film Hall mobility and electron concentration. From Hall measurements, it was found that films grown at TMGa flows >200 μmol/min. had lower mobilities and electron concentrations than those grown at <200 μmol/min., implying a compensation mechanism at lower VI/III ratios ($O_2$ flow was constant for all growths). This agrees with other literature findings that at lower VI/III ratios, the amount of Carbon in the films increases, which can lead to electron compensation[18,54]. Thus, for all subsequent growths, the TMGa flow was kept at an optimal rate of 170 μmol/min. to maximize Hall mobility and growth rate while minimizing Carbon impurities. Figure 2(a) shows a plot of the Hall mobility vs. Hall electron concentration for a set of films all grown at 170 μmol/min. In this figure, the hollow shapes denote intentional doping with Silane, whereas the filled shapes denote UID films. The UID electron concentration was found to consistently be between 2-4 x $10^{15}$ cm$^{-3}$ for all growths, but the mobility was found to be inversely related to the growth pressure, with lower pressures giving consistently higher mobilities. This phenomenon is potentially due to the reduced gas-phase pre-reactions at lower pressures, so less pre-reacted adducts would find their way to the growth surface, hence improving the film quality and Hall electron mobility at lower pressures. At 15 Torr, Hall mobility values were all greater than 150 cm$^2$/Vs with a maximum of 176 cm$^2$/Vs reached, signifying good crystalline quality and minimal impurities in the TMGa drift layers. In addition, the effect of TMGa flow and pressure on RMS roughness was studied. Surface roughness values were measured via atomic force microscopy (AFM) scans on 2 x 2 μm$^2$ surface areas. Figure 2(b) shows that the RMS roughness values significantly increase with increased TMGa flow at 60 Torr, going from 5 nm to 32 nm RMS. However, this trend was not observed at lower pressures, with films maintaining similar RMS roughness values of ~5 nm across all observed TMGa flow rates at 15 Torr. This could be attributed to an increased surface diffusion length at lower pressures[60] leading to favorable growth modes and suppression of 2D island growth.

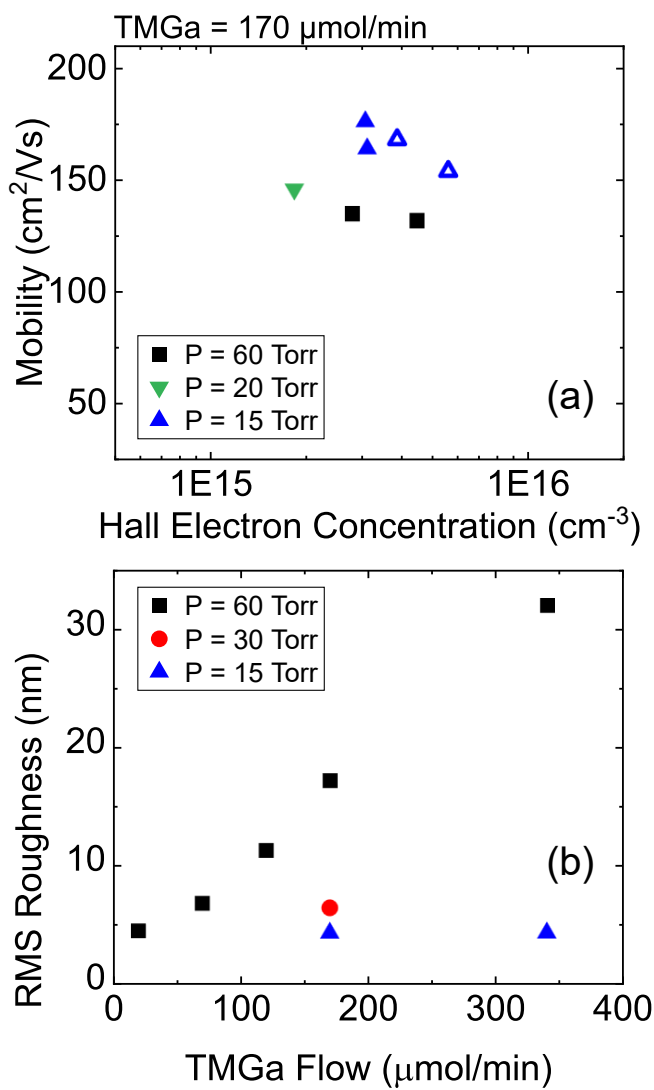

FIG. 2. (a) Plot of electron Hall mobility vs. Hall electron concentration from 4-corner Van Der Pauw Hall structures. Hollow shapes indicate intentionally doped films, and solid shapes are UID films. (b) Dependence of RMS roughness on TMGa Flow rate and growth chamber pressure.

## C. Secondary ion mass spectroscopy

To further analyze the concentration of impurities in the thick TMGa epilayers, secondary ion mass spectroscopy (SIMS) was performed courtesy of EAG laboratories. The film analyzed was grown using the UID conditions that gave the best mobilities in Figure 2(a), with a TMGa flow of 170 μmol/min., temperature of 1000 °C, pressure of 15 Torr, and VI/III ratio of 523. This condition produced a film with a UID Hall electron concentration of ~3 x $10^{15}$ cm⁻³. Figure 3 depicts the SIMS scan of a 450 μm film grown at the same conditions. In literature, unintentional Si impurities were found to likely be the source of the UID electron concentration in thick drift layers[23]. Since the SIMS silicon concentration was found to be at or below the tool detection limit ($2 \times 10^{15}$ cm$^{-3}$), which is comparable to the electron concentration from Hall measurements ($3 \times 10^{15}$ cm$^{-3}$), it is clear there is negligible electron compensation in the TMGa epilayers. In addition, the C and H atomic concentrations in Figure 3 were found to be at or below the detection limit, which is consistent with the above analysis that the films have low concentrations of unintentional impurities.

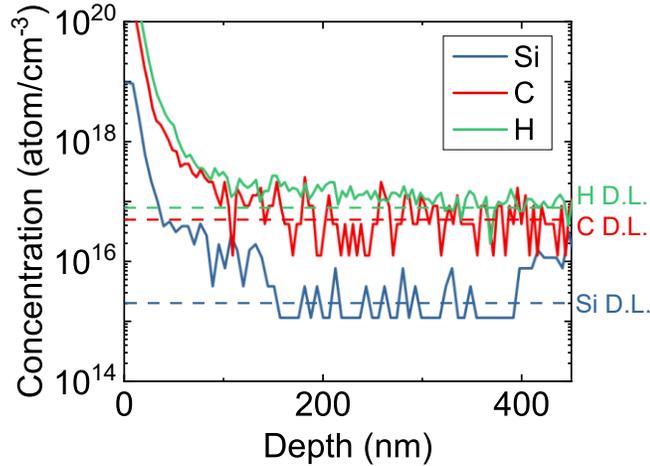

FIG. 3. SIMS scan of a representative UID MOCVD-grown TMGa epilayer, showing a Si, C, and H level at or below the detection limit, indicating low unintentional impurity concentrations in the films.

## D. Diode epilayer growth

The optimal growth conditions for thick epilayers were found to be at a TMGa flow of 170 μmol/min., O$_2$ flow of 0.089 mol/min., VI/III ratio of 523, growth rate of 4 μm/hr., temperature of 1000 °C, and a growth pressure of 15 Torr. With these growth conditions, we grew a 10 μm thick intentionally doped film on co-loaded HF treated Fe-doped semi-insulating and non-HF treated Sn-doped conductive (010) β-Ga$_2$O$_3$ substrates. This film was intentionally doped using diluted Silane in N$_2$, with a molar flow rate of 23.9 pmol/min. The Fe-doped substrate was processed into a 4-corner Van Der Pauw Hall structure and Hall measurements were performed. Results gave a Hall mobility of 130 cm$^2$/Vs, electron concentration of $5.57 \times 10^{15}$ cm$^{-3}$, and sheet resistance of 8.6 kΩ/□. The as-grown surface analysis was done via AFM, differential interface contrast (DIC) microscopy, and SEM. The AFM in Figure 4(a) shows the standard surface morphology for homoepitaxially grown (010) films, with extended islands/grooves along the enhanced adatom mobility [001] direction[12,14,20,23,27,62–64]. The RMS roughness of the film was found to be 7.46 nm. The slight increase in surface roughness from the previous ~5 nm RMS at this growth condition in Figure 2(b) is potentially due to the 10 μm film being ~60% thicker. Figure 4(b) shows a large area surface scan using DIC optical microscopy and shows the presence of many surface defects. The microstructure of these defects is in the form of hillocks extending in the [001] direction, as seen in the SEM image in Figure 4(c). These defects are the same microstructure observed in our previous work using TEGa precursor[23], meaning they do not depend on the Ga precursor used. The main types of defects seen in MOCVD growths are inverted pyramidal pits[65,66], 3D pyramids[22], spherical particles[13], and hillocks oriented in the [001] direction[20,23,65]. In this work, only hillock defects were found to be present on the epilayer surface, with the density of hillocks increasing with both longer growth times and larger epilayer thickness. Since $\beta$-Ga$_2$O$_3$ particles can form in the gas phase due to pre-reactions before reaching the growth interface[13,67], it is possible that these particles can deposit on the growth surface and interrupt the growth locally, creating an extended crystalline defect structure. With an increase in growth thickness or growth time, there will be an increase in the amount of TMGa entering the reactor and thus an increase in the number of pre-reaction induced particle depositions that occur during growth, explaining why the number of defects increases with growth thickness or growth time. Additionally, it has been shown that hillock structures can also arise from twin defects originating from a surface facet on an imperfect substrate/epitaxial layer interface[65]. It is most likely that there is a combination of both particle and substrate/epitaxial interface mediated defect formation during growth. Thus, these defect nucleation sources need to be mitigated going forward to reduce the defect density on (010) drift layers. From literature, potential methods for accomplishing lower defect densities include using vicinal substrates[22] and reducing gas-phase pre-reactions by decreasing the distance between the MOCVD showerhead and susceptor[68].

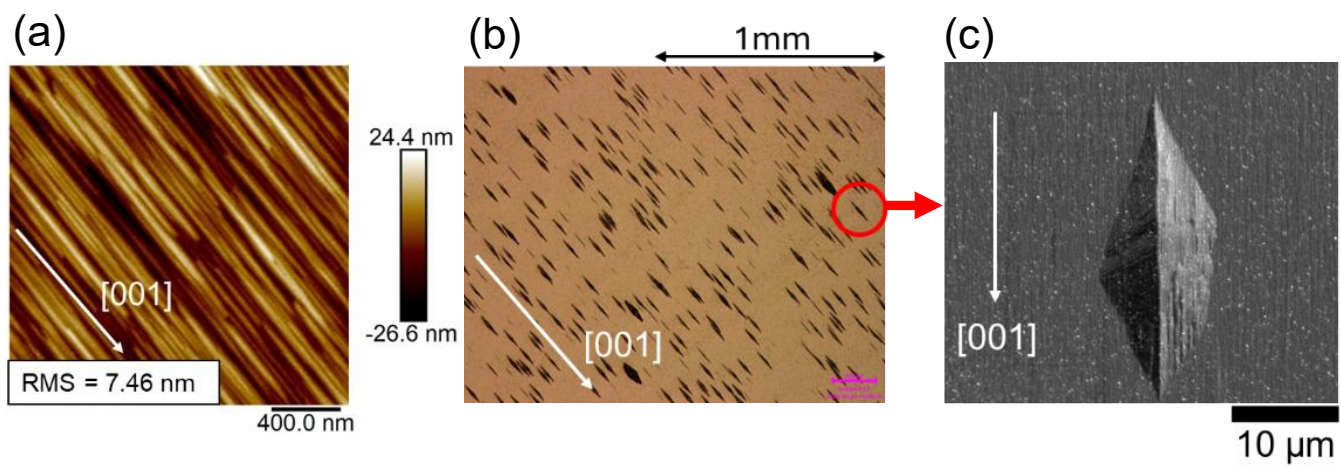

FIG. 4. (a) AFM scan of the 10 μm thick intentionally doped epilayer, showing an RMS roughness of 7.46 nm (b) differential interface contrast (DIC) microscopy scan of the sample surface, showing hillock shaped defects oriented along the [001] direction (c) enlarged SEM image of a single hillock defect.

# III. DIODE RESULTS

## A. Fabrication

After growth and surface characterization, field plated Schottky barrier diodes (FP-SBDs) were fabricated on the 10 μm epilayer. Device fabrication began with the patterning and deposition of a Ni/Au/Ni Anode metal using optical lithography and e-beam evaporation with a lift-off process. The Schottky anode metal was deposited first in the process to protect the epitaxial layer from any ion damage that could be induced by future plasma etching. After the anode metal, a 120 nm $TiO_2/Al_2O_3$ nanolaminate dielectric was deposited using a plasma atomic layer deposition (ALD) process at 300 °C. This dielectric had a dielectric constant of 17 and was capped on both the bottom and top with 5nm of $Al_2O_3$ to improve the leakage performance[69]. After ALD, the dielectric above the anode metal was removed using a $BCl_3$ inductively coupled plasma (ICP) etch. This etch was then followed by the field plate metal deposition, which consisted of another e-beam deposited Ni/Au metal stack. This top metal stack connected to the bottom anode metal and was lithographically patterned to overlap the dielectric by 10 μm, creating the field plate structure seen in Figure 5(a). Finally, a Ti/Au metal stack was deposited on the backside of the Sn-doped substrate to create the ohmic cathode contact.

## B. High voltage CV

To electrically verify the thickness and net donor concentration of the epilayer electrically, large area 1 mm² MOS diodes were measured. High voltage C-V measurements were performed on a Keysight B1505a parameter analyzer to extract the net donor concentration ($N_D$-$N_A$) vs. depth profile which is seen in Figure 5(b). The profile is mostly flat, with the net donor concentration increasing from 4 x $10^{15}$ to 8 x $10^{15}$ $cm^{-3}$ towards the substrate. Potential reasons for the increase in doping include surface riding of silicon contamination at the substrate/epitaxial interface into the film during growth at high temperatures[62] or diffusion of Sn from the substrate[23]. The charge in the epitaxial layer could not be fully depleted due to MOS leakage and premature breakdown of the measurement pads, explaining why the $N_D$-$N_A$ profile in Figure 5(b) becomes noisier and truncates before reaching the substrate/epitaxial layer interface.

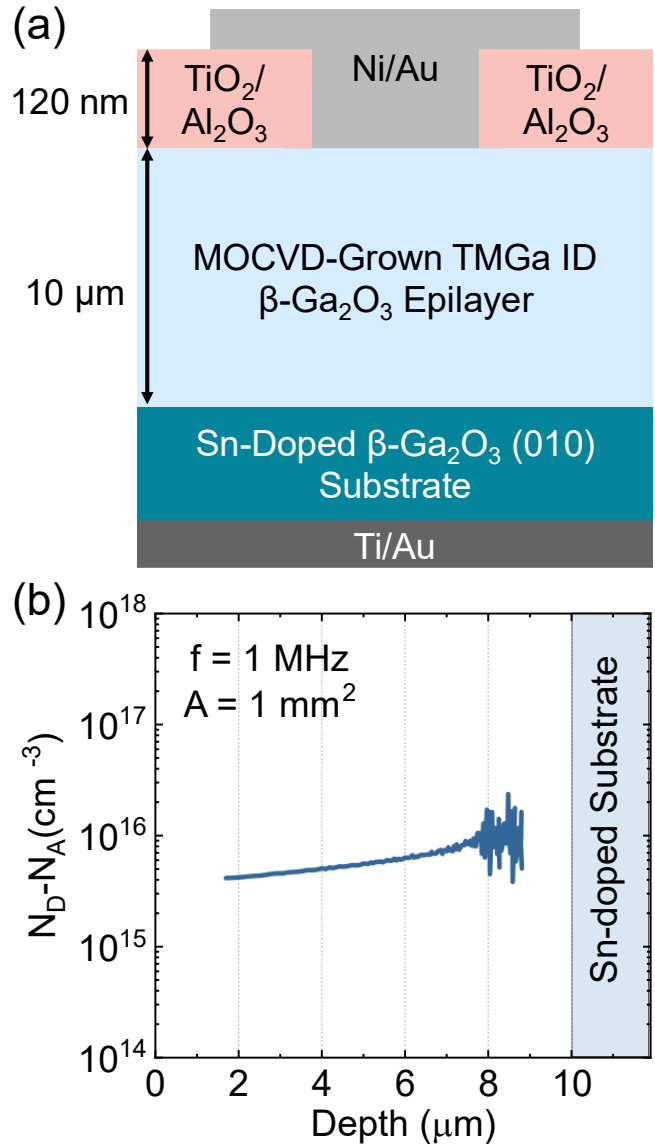

FIG. 5. (a) Device structure of the field plated Schottky barrier diode fabricated on the 10 μm thick MOCVD epilayer (b) $N_D$-$N_A$ for the 10 μm thick epilayer extracted from high voltage MOS C-V measurements. C-V structure broke down prematurely, so the entire 10 μm could not be measured.

## C. J-V Characteristics

The current density vs. voltage (J-V) measurements were performed on a Keithley 4200 parameter analyzer to determine the on and off-state performance of the FP-SBDs. Current density was normalized with respect to the anode contact diameter with a current spreading angle of 45° considered. Figure 6(a) is a plot of the linear scale J-V, showing standard Schottky diode characteristics with a turn on voltage of 1 V and a high current density of >100 A/cm² at 3V forward bias. The differential specific on resistance ($R_{on,sp}$) was also measured and plotted in Figure 6(a), with a minimum $R_{on,sp}$ value of 16.22

mΩ.cm². If the epilayer's drift mobility is assumed to be slightly lower than the Hall mobility due to non-unity Hall factor, as was found in our previous work on low-doped epilayers[12], the drift layer's contribution to the total $R_{on,sp}$ can be estimated. Thus, using an estimated drift mobility of 110 cm²/Vs and the 45° current spreading model for a planar SBD[70], the total $R_{on,sp}$ for the drift layer was calculated to be 12.38 mΩ.cm². Based on previous work[12], the maximum contribution of the substrate resistance to the overall $R_{on,sp}$ is ~0.8 mΩ.cm². Thus, the theoretically calculated total $R_{on,sp}$ = 13.18 mΩ.cm², whereas the measured value was 16.22 mΩ.cm². The discrepancy between the calculated and measured values is likely due mostly to poor Ohmic contact to the backside of the wafer, as seen in Figure 6(c). It is also possible that a lower epitaxial drift mobility or higher substrate resistance could play a role. The log-scale J-V in Figure 6(b) shows a high diode on-off current ratio of ~$10^{10}$ and a near unity ideality factor of 1.04, demonstrating highly rectifying behavior and a high-quality metal/semiconductor Schottky interface.

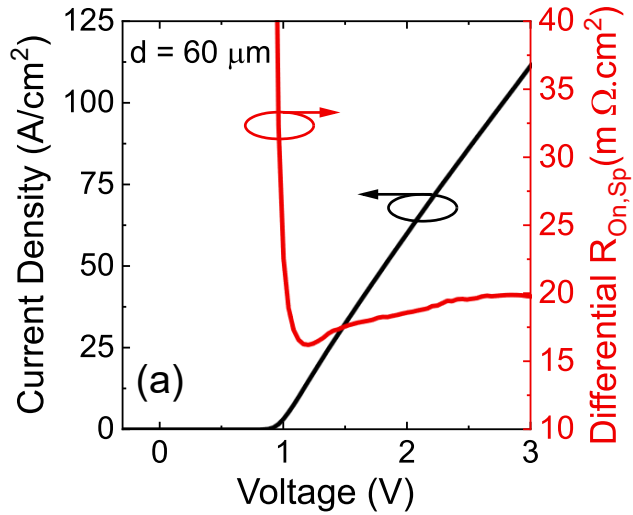

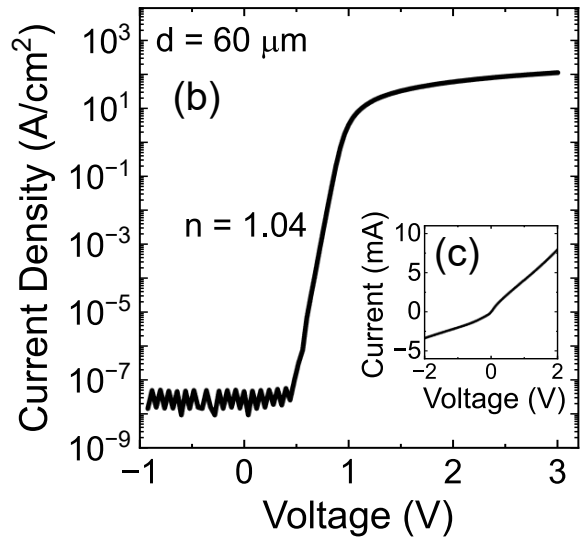

FIG. 6. (a) Current density and differential on resistance vs. voltage plot for a 60 μm diameter FP-SBD showing appreciable on-current and a minimum $R_{on,sp}$ of 16.22 mΩ.cm² (b) log-scale J-V characteristics, showing a ~$10^{10}$ rectification ratio and a near unity ideality factor of 1.04 (c) I-V of the backside Ohmic contact, showing non-linear behavior

## D. Device breakdown

High voltage breakdown measurements were performed on a Keysight B1505a parameter analyzer using Fluorinert as a dielectric fluid to prevent air breakdown induced arcing during measurement. The breakdown voltage ($V_{BR}$) of 60 μm diameter diodes was found to be 1.50 kV. To determine the effectiveness of the field plate's ability to reduce the peak electric field, planar SBDs were also measured alongside the FP-SBDs. The results plotted in Figure 7(a) show the planar Schottky structures began to leak at a much lower voltage (50-100 V) compared to the FP-SBDs (500 V). This means the FP structure successfully spread the electric field across the FP oxide, reducing the peak field at the anode corner and delaying the onset of field-mediated leakage current through the Schottky barrier[71,72]. This earlier leakage in the planar SBDs led to premature breakdown when compared to the FP-SBDs, with the planar SBDs only reaching a $V_{BR}$ of 1 kV. Thus, adding the field plate improved the breakdown by ~500 V. The parallel plane electric field at breakdown under the anode center ($E_{∥,max}$) was calculated using a punch-through diode model and was found to be 2.04 MV/cm and 2.74 MV/cm for the FP-SBD and MOS diodes respectively, which is a state-of-the-art $E_{∥,max}$ result for any diode created on MOCVD-grown thick (010) epilayers and highlights the high quality of the grown epilayer.

Since the breakdown nature of β-Ga₂O₃ unipolar devices is catastrophic in nature, there is material damage left behind after a high-voltage breakdown measurement. In these films, a consistent crystallographic breakdown was observed along the [001] direction, with breakdown occurring at a hillock defect if one was present in the device footprint, as seen in Figure 7(b). Since the trench formed after breakdown was consistently located in the exact position of where a defect used to be, hillock defects are considered killer defects when it comes to electrical breakdown performance and must be mitigated for future high voltage devices. The observation of trenches forming along the [001] orientation after breakdown suggests some form of crystallographic failure of the material. In literature, fin diodes fabricated on (001) β-Ga₂O₃ substrates exhibited cracking along the [010] direction after destructive breakdown. This cracking was suggested to be due to thermo-mechanical failure of the (100) cleavage planes due to heat generation during breakdown[73]. In the case of the current MOCVD FP-SBDs fabricated on (010) substrates, the observed trenches in the [001] direction also have (100) oriented sidewalls, implying that the breakdown features are also due to thermo-mechanically induced cracking of the (100) cleavage planes.

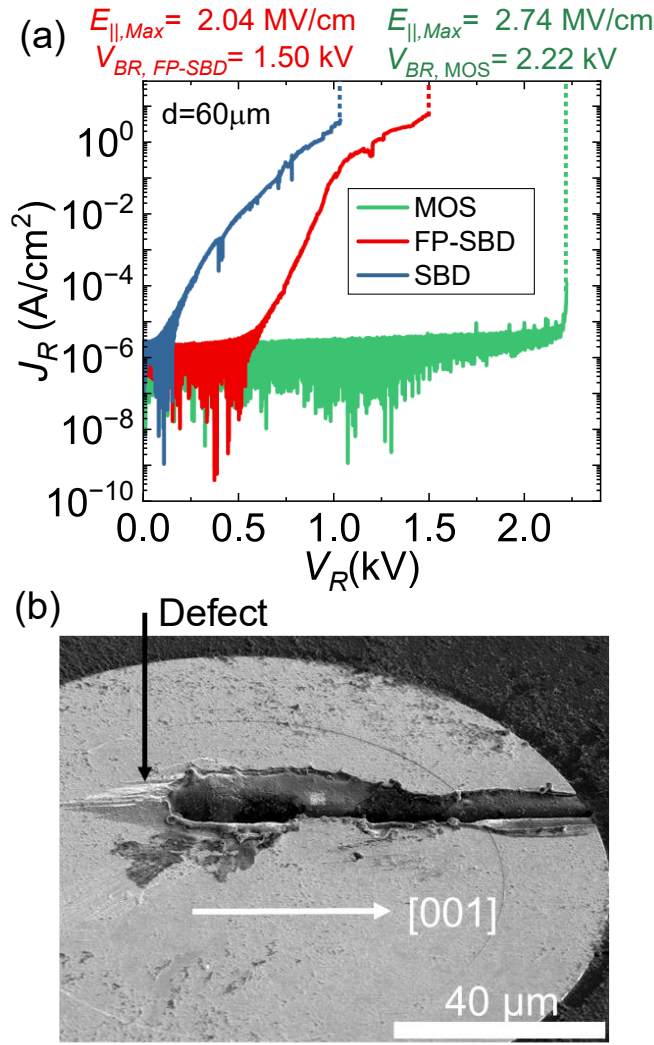

FIG. 7. (a) Current density vs. reverse bias voltage breakdown plot showing a breakdown of 1.50 kV for FP-SBDs and 2.22 kV for MOS diodes with a state-of-the art parallel plane electric field of 2.04 MV/cm and 2.74 MV/cm for FP-SBD and MOS respectively at breakdown. (b) Defect mediated destructive breakdown showing that breakdown occurs at a defect and cracks along the [001] direction, indicating potential thermo-mechanically induced cracking of the (100) cleavage plane.

## IV. CONCLUSIONS

Growth of up to 10 µm thick intentionally doped β-Ga$_2$O$_3$ drift layers with electron concentrations as low as 3 x 10$^{15}$ cm$^{-3}$ have been demonstrated with fabrication of kilovolt-class field plated Schottky barrier diodes on these thick drift layers. Growth parameter sweeps found that once carbon compensation was minimized via high growth temperatures and high VI/III ratios, growth pressure was the most impactful variable for achieving high quality thick epilayers. Decreasing the pressure improved the growth rate, mobility, and RMS surface roughness of the samples. Growth rates of up to 7.2 µm/hr., thicknesses of up to 10µm, Hall mobilities of up to 176 $cm^2$/Vs, RMS roughness down to 5.45 nm, UID concentrations as low as 2 x $10^{15}$ $cm^{-3}$, and controllable intentional doping down to 3 x $10^{15}$ $cm^{-3}$ were achieved for the thick MOCVD epilayers. After growth optimization, FP-SBDs were fabricated on a 6.5 x $10^{15}$ $cm^{-3}$ intentionally doped 10 µm thick film. The FP-SBDs were found to have current densities >100 A/$cm^2$, a specific differential on resistance ($R_{on,sp}$) of 16.22 mΩ.$cm^2$, an ideality factor of 1.04, and a maximum breakdown voltage of 1.50 kV, leading to a punch-through maximum field of 2.04 MV/cm under the Anode metal. The fabricated FP-SBDs achieved state-of-the-art results for any SBDs on MOCVD-grown (010) drift layers, further highlighting MOCVD as a promising growth method for thick drift layers capable of supporting multi-kV class diodes and transistors.

# ACKNOWLEDGMENTS

The authors acknowledge funding from the ARPA-E ULTRAFAST program (DE-AR0001824) and Coherent / II-VI Foundation Block Gift Program. A portion of this work was performed at the UCSB Nanofabrication Facility, an open access laboratory.

# AUTHOR DECLARATIONS

The authors have no conflict to disclose.

# DATA AVAILABILITY

The data that support the findings of this study are available from the corresponding author upon reasonable request.